\documentclass[12pt]{article}
\usepackage{latexsym}
\usepackage{psfig,amssymb,epsf}
\def\bseq{\begin{subequation}}  
\def\eseq{\end{subequation}}
\def\bsea{\begin{subeqnarray}}  
\def\esea{\end{subeqnarray}}

\newcommand{\bbox}{\lower.2ex\hbox{$\Box$}}

\newcommand{\beq}{\begin{equation}}
\newcommand{\eeq}{\end{equation}}
\newcommand{\bea}{\begin{eqnarray}}
\newcommand{\eea}{\end{eqnarray}}
\newcommand{\ena}{\end{eqnarray}}
\newcommand{\ba}{\begin{array}}
\newcommand{\ea}{\end{array}}
\newcommand{\ben}{\begin{enumerate}}
\newcommand{\een}{\end{enumerate}}
\newcommand{\bde}{\begin{description}}
\newcommand{\ede}{\end{description}}

\renewcommand{\a}{\alpha}

\newcommand{\pa}{\partial}
\newcommand{\g}{\gamma}

\newcommand{\D}{\Delta}
\newcommand{\e}{\epsilon}

\newcommand{\p}{\pi}

\newcommand{\ad}{{\dot{\alpha}}}

\begin{document}
\begin{titlepage}
\begin{flushright}
IFUM-718-FT\\

\end{flushright}
\vspace{2cm}
\noindent{\Large \bf Exact anomalous dimensions of ${\cal N}=4$ Yang-Mills  }\\
\vskip 0.2mm
\noindent{\Large \bf operators with large $R$ charge}
\vspace{1cm}
{\bf \hrule width 16.cm}
\vspace {1cm}
\begin{center}
{\large \bf  Alberto Santambrogio }\\
{\small \bf and}\\
{\large \bf Daniela Zanon}

\vskip 2mm
{ \small

\noindent Dipartimento di Fisica dell'Universit\`a di Milano
and

\noindent INFN, Sezione di Milano, Via Celoria 16,
20133 Milano, Italy}
\end{center}
\vfill
\begin{center}
{\bf Abstract}
\end{center}
{\small In a ${\cal N}=1$ superspace formulation of ${\cal N}=4$ Yang-Mills theory we obtain the anomalous dimensions of chiral operators with large $R$ charge $J \rightarrow \infty$ keeping $g^2 N/J^2$ finite, to all orders of perturbation theory in the planar limit. Our result proves the conjecture that the anomalous dimensions are indeed finite in the above limit. This amounts to an exact check of the proposed duality between a sector of ${\cal N}=4$ Yang-Mills theory with large $R$ charge $J$ and string theory in a pp-wave background.}
\vspace{2mm}
\vfill \hrule width 6.cm
\begin{flushleft}
e-mail: alberto.santambrogio@mi.infn.it\\
e-mail: daniela.zanon@mi.infn.it
\end{flushleft}
\end{titlepage}

Recently the existence of a new  duality \cite{BMN} has attracted much attention: it has been argued that the  $IIB$ string on a pp-wave background \cite{M,BFHP} is dual to the sector of ${\cal N}=4$ supersymmetric $SU(N)$ Yang-Mills theory with large $R$ charge $J$, in the limit $N,J \rightarrow \infty$ and $g^2 N/J^2$ finite. More precisely  a certain correspondence has been suggested between string states with light cone momentum $p^+$ and energy $p^-$, and operators in the Yang-Mills theory with conformal dimensions $\D$ and charge $J$, related by \cite{BMN}
\beq
2p^-=\mu(\D-J) \qquad\qquad 2p^+=\frac{\D+J}{\mu R^2}\qquad\qquad 
R^2= 2\a' \sqrt{\pi g_s N}
\eeq
In particular according to the above correspondence the mass spectrum of string states should match the anomalous dimensions of dual operators in the field theory. In the planar limit these operators are multiplicatively renormalized \cite{KPSS,GMR,CFHMMPS}.

In this paper we obtain to all orders in perturbation theory the anomalous dimensions of this class of operators, confirming the proposals presented in \cite{BMN} and in \cite{GMR}.  
Our derivation is based on the constraints imposed on the two-point functions by the superconformal invariance of the theory. 
The analysis is performed using ${\cal N}=1$ superspace techniques that are very efficient for correlator computations in supersymmetric theories. A perturbative calculation of the two-point functions of the CPO operators at ${\cal{O}}(g^4)$ was performed in ${\cal N}=1$ superspace in \cite{noi}. 

\vspace{0.6cm}
The ${\cal N}=4$ supersymmetric Yang-Mills classical action
written in terms of ${\cal N}=1$ superfields (we use the
notations and conventions adopted in \cite{superspace}) is given by
\bea
&&S= {\rm Tr} \left( \int~ d^4x~d^4\theta~ e^{-gV}
\bar{\Phi}_i e^{gV} \Phi^i +\frac{1}{2g^2}~ \left[\int ~d^4x~d^2\theta~ W^\a W_\a + \int ~d^4x~d^2\bar{\theta}~ \bar{W}^\ad \bar{W}_\ad\right]\right.\nonumber\\
&&\left.~~~~~~~~~~~~~+\frac{ig}{3!} \int ~d^4x~d^2\theta~ \e_{ijk}
 \Phi^i
[\Phi^j,\Phi^k] - \frac{ig}{3!}\int ~d^4x~d^2\bar{\theta}~ i\e^{ijk} \bar{\Phi}_i
[\bar{\Phi}_j,\bar{\Phi}_k] \right)
\label{N4SYMaction}
\eea
where the $\Phi^i$ with $i=1,2,3$ are three chiral
superfields, and the $W^\a= i\bar{D}^2(e^{-gV}D^\a e^{gV})$ are the gauge
superfield strengths. All the fields are Lie-algebra valued, e.g.
$\Phi^i=\Phi^i_a T^a$, in the adjoint representation of the gauge group $SU(N)$.
The chiral superfields $\Phi^i$ contain the six real scalar fields ${\cal{X}}^1,\ldots,{\cal{X}}^6$ of the ${\cal N}=4$ Yang-Mills theory.
For notational simplicity it is convenient to rename the three chiral superfields
 as $\Phi={\cal{X}}^1+i {\cal{X}}^2$, $\Psi={\cal{X}}^3+i {\cal{X}}^4$, $Z={\cal X}^5+i{\cal X}^6$. With this notation the chiral interaction vertices can be written as
\beq
{\rm Tr}\left(ig\int d^2 \theta (\Phi Z\Psi -\Phi \Psi Z) +~h.c.\right)
\label{chiralvertex}
\eeq
We consider the $U(1)$ subgroup of the $SU(4)$ $R$-symmetry group which rotates ${\cal X}^5$ and ${\cal X}^6$ and leaves the other four ${\cal X}^i$ invariant, so that $Z$ has unit charge with respect to this subgroup while $\Phi$ and $\Psi$ are neutral.

Now we want to consider operators with a large number of charged $Z$ fields, corresponding to large $J$ operators, with insertions of a small number of neutral fields. An example of such an operator with two insertions is
\beq
\sum_l e^{il\varphi}{\rm{Tr}}\left({\cal X}^1 Z^l {\cal X}^3 Z^{J-l}\right)
\label{example}
\eeq
where we have defined
\beq
\varphi=\frac{2\p n}{J}
\label{defphi}
\eeq
with $n$ the excitation level of the dual string state.
Under the assumption of a ``dilute gas" approximation \cite{BMN}
we focus on the insertion of one neutral field. 
Following the work in \cite{GMR} we consider the operators
\beq
\sum_l e^{il\varphi} Z^l {\cal X}^i Z^{J-l}
\label{building}
\eeq
with $i=1,\ldots,4$ as ``building blocks" for operators with more insertions.
More precisely we concentrate on the following two operators
\beq
{\cal{O}}_J= \sum_l e^{il\varphi} Z^l\Phi Z^{J-l}
\label{operatorO}
\eeq
and \footnote{Gauge invariance would require to study the operator ${\cal{U}}_J=\sum e^{il\varphi} Z^l e^{-gV}\bar{\Psi}e^{gV}Z^{J-l}$ and use gauge covariant derivatives.
However for our purposes it is sufficient to work in a flat background.}
\beq
{\cal{U}}_J=\sum_l e^{il\varphi} Z^l \bar{\Psi}Z^{J-l}
\label{operatorU}
\eeq
The suffix $J$ counts the number of charged $Z$-fields, while $\varphi$ is always defined as in (\ref{defphi}).
From the point of view of the ${\cal N}=4$ theory they  belong to the same multiplet and so they have the same properties: 
they renormalize in exactly the same way and in particular they have the same anomalous dimension $\g$. This is the quantity that we want to determine, in 
the planar limit and to all orders in perturbation theory.

To this end, inspired by similar techniques used in \cite{A},
we make use of the equations of motion from the action in (\ref{N4SYMaction})
\beq
\bar{D}^2\bar{\Psi}=ig(\Phi Z-Z\Phi)\qquad\qquad D^2\Psi=-i g(\bar{\Phi}\bar{Z}-\bar{Z}\bar{\Phi})
\label{eqmotion}
\eeq
and relate the ${\cal{U}}_J$ operators to the ${\cal{O}}_J$ ones. Indeed
we obtain
\bea
&&\bar{D}^2{\cal{U}}_J=-i g(e^{-i\varphi}-1){\cal{O}}_{J+1}\nonumber\\
&&D^2\bar{{\cal{U}}}_J=ig(e^{i\varphi}-1)\bar{{\cal{O}}}_{J+1}
\label{mainrel}
\eea

The equations in (\ref{mainrel}) allow to express the two-point correlation functions of the ${\cal{U}}_J$ operators in terms of the ${\cal{O}}_J$'s.
With the definition
\beq
\a=e^{i\varphi}+e^{-i\varphi}-2=-(e^{i\varphi}-1)(e^{-i\varphi}-1)
\label{def}
\eeq
we obtain ($z\equiv x,\theta$)
\beq
<\bar{D}^2{\cal{U}}_J(z)D^2\bar{{\cal{U}}}_J(z')>=-g^2 \a <{\cal{O}}_{J+1}(z)
\bar{{\cal{O}}}_{J+1}(z')>
\label{relcorr}
\eeq
We notice that ${\cal{O}}_{J+1}$ differs from ${\cal{O}}_{J}$ only through the presence of an extra $Z$-field. In the planar limit and in the ``dilute gas" approximation, it is easy to see that only a finite number of neighbouring $Z$-fields interact with the impurity at a given order in perturbation theory \cite{GMR}. Thus the extra $Z$-field present in ${\cal{O}}_{J+1}$ does not play any relevant role and the renormalization of ${\cal{O}}_{J+1}$ is the same as the one of ${\cal{O}}_{J}$.
Now, using the above relation and the fact that ${\cal{U}}_J$ and ${\cal{O}}_{J+1}$ have the same renormalization properties, we want to compute their anomalous dimension $\g$.

First we find a general expression for the two-point correlator of an ${\cal{N}}=4$
operator ${\cal{A}}_{(h,\bar{h})}$ which contains $h$ chiral and $\bar{h}$ antichiral superfields.
In ${\cal{N}}=1$ superspace it is rather simple to compute explicitly such a two-point function  at the tree-level. One has propagators given by
\\
\beq
<\Phi(z)\bar{\Phi}(z')> =\frac{1}{4\p^2}\bar{D}^2\frac{\delta^4(\theta-\theta')}{|x-x'|^{2}}\overleftarrow{D}^2
\label{prop}
\eeq
\\
and performs straightforward $D$-algebra. 
Then one finds
\bea
&&<{\cal{A}}_{(h,\bar{h})}(z)\bar{{\cal{A}}}_{(h,\bar{h})}(z')>_{tree}=
<(\Phi^h\bar{\Phi}^{\bar{h}})(z)(\bar{\Phi}^h\Phi^{\bar{h}})(z')>_{tree}=
\nonumber\\
&&~~~~~\nonumber\\
&&~~~~=f_{{\cal{A}}}
\left\{\frac{1}{2}D^\a \bar D^2 D_\a + \frac{w}{4\Delta}
[ D^\a,\bar D^{\ad} ] i\pa_{\a\ad} + 
\frac{\Delta^2 + w^2 -2\Delta}{4\Delta (\Delta - 1)}\square\right\}
\frac{\delta^4(\theta-\theta')}{|x-x'|^{2\Delta}}
\label{niceformula}
\eea
where $\Delta = h + \bar{h}$ is the total dimension and $w = h-\bar{h}$
is the chiral weight. We have introduced a proportionality factor $f_{{\cal{A}}}$ which at tree level can be easily computed. 

Now we take advantage from the fact that the ${\cal N}=4$ supersymmetric Yang-Mills theory is
conformally invariant. In such a theory two-point correlation functions are determined by their conformal dimensions \cite{PO} and this allows us to immediately extend (\ref{niceformula}) to the complete interacting theory. Being $\g$  the anomalous dimension of the operator we simply have to substitute
\\
\beq
\D\rightarrow \D+\g
\label{anomalous}
\eeq
\\
while keeping $w$ unchanged. In fact the chiral weight $w$ is not renormalized because of ${\cal N}=4$ supersymmetry. Thus we obtain
\bea
&&<{\cal{A}}_{(h,\bar{h})}(z)\bar{{\cal{A}}}_{(h,\bar{h})}(z')>
=f_{{\cal{A}}}(g^2,N,h,\bar{h})
\left\{\frac{1}{2}D^\a \bar D^2 D_\a +  \frac{w}{4(\Delta+\g)}
[ D^\a,\bar D^{\ad} ] i\pa_{\a\ad} \right.~~~~~~\nonumber\\
&&~~~~~\nonumber\\
 &&~~~~~~~~~~~~~~~~~~~~~~~~~~~~~~~\left.+ 
\frac{(\Delta+\g)^2 + w^2 -2(\Delta+\g)}{4(\Delta+\g) (\Delta +\g- 1)}\square\right\}
\frac{\delta^4(\theta-\theta')}{|x-x'|^{2(\Delta+\g)}}
\label{niceformula2}
\eea
\\
where the $f_{{\cal{A}}}$ is a function not fixed by conformal invariance. In the planar limit we are considering, the dependence on $g$ and $N$ is through the 't Hooft coupling $g^2N$.

The expression in (\ref{niceformula2}) can be specialized to the two-point functions of the ${\cal{U}}_{J}$ and ${\cal{O}}_{J+1}$ operators. For the former we have
$\D=J+1$, $w=J-1$,  for the latter  $\D=J+2$, $w=J+2$.
Using the above values for the dimensions $\D$ and weights $w$ we easily 
obtain the correlator for the ${\cal{U}}_{J}$'s
\bea
&&<\bar{D}^2{\cal{U}}_J(z)D^2\bar{{\cal{U}}}_J(z')>=\nonumber\\
&&~~~~~~\nonumber\\
&&~~~=\frac{N^{J+1}}{(4\p^2)^{J+1}}
f(g^2N,J)\bar{D}^2\left\{\frac{1}{2}D^\a \bar D^2 D_\a + \frac{J-1}{4(J+1+\g)}
[ D^\a,\bar D^{\ad} ] i\pa_{\a\ad} \right.~~~~~~\nonumber\\
&&~~~~~\nonumber\\
 &&~~~~~\left.+ 
\frac{(J+1+\g)^2 + (J-1)^2 -2(J+1+\g)}{4(J+1+\g) (J +\g)}\square\right\}
D^2\frac{\delta^4(\theta-\theta')}{|x-x'|^{2(J+1+\g)}}\nonumber\\
&&~~~= \frac{N^{J+1}}{(4\p^2)^{J+1}}
f(g^2N,J) (\g^2+2\g)\bar{D}^2 D^2\frac{\delta^4(\theta-\theta')}
{|x-x'|^{2(J+2+\g)}}
\label{corrUU}
\eea
In the same way we find
\bea
&&<{\cal{O}}_{J+1}(z)
\bar{{\cal{O}}}_{J+1}(z')>=\frac{N^{J+2}}{(4\p^2)^{J+2}}
f(g^2N,J)\Big(\bar{D}^2D^2
+~~~~~~~~~~~\nonumber\\
&&\left.-\frac{\g}{4(J+2+\g)}[ D^\a,\bar D^{\ad} ] i\pa_{\a\ad} -
\frac{\g(2J+4+\g)}{(J+2+\g) (J+1 +\g)}\square\right)
\frac{\delta^4(\theta-\theta')}{|x-x'|^{2(J+2+\g)}}
\label{corrOO}
\eea
In (\ref{corrUU}) and (\ref{corrOO}) we have inserted the tree level constants so that
the common function $f(g^2N,J)$ is of the form $f(g^2N,J)=1+ g^2 Nf_1(J)+\ldots$. At this point we substitute the above expressions in (\ref{relcorr}) and obtain
\bea
&&\frac{N^{J+1}}{(4\p^2)^{J+1}}
f(g^2N,J) (\g^2+2\g)\bar{D}^2 D^2\frac{\delta^4(\theta-\theta')}{|x-x'|^{2(J+2+\g)}}
=\nonumber\\
&&~~~~~~~~~=-g^2 \a\frac{N^{J+2}}{(4\p^2)^{J+2}}
f(g^2N,J)\bar{D}^2 D^2\frac{\delta^4(\theta-\theta')}{|x-x'|^{2(J+2+\g)}}
\label{inter}
\eea
where we have dropped terms from the r.h.s. of (\ref{corrOO}) which would produce subleading contributions to $\g$ in the large $J$ limit.
Thus we are lead to a very simple equation which determines the anomalous dimension
\beq
\g^2+2\g=- \frac{g^2 N}{4\p^2}\a
\label{final}
\eeq
The solution is simply given by
\beq
\g=-1+\sqrt{1-\frac{g^2 N}{4\p^2} \a}
\label{anomalous1}
\eeq
(The solution with the minus sign in front of the square root  would lead to  $\g=-2$  at the tree level.)
In order to be consistent with the large $J$ approximation used so far, we have to  truncate
$\a\sim -\varphi^2$, so that we finally obtain
\beq
\g=-1+\sqrt{1+\frac{ g^2 N n^2}{J^2}}
\label{anomalousvera}
\eeq

We emphasize that the result in (\ref{anomalousvera}) gives the anomalous dimension of the operators ${\cal{O}}_J$ to all orders in perturbation theory. It confirms the conjecture given in \cite{GMR} and exactly coincides with the value obtained at strong coupling from the dual string approach \cite{BMN}.
This amounts to an exact check of the proposed duality between a sector of ${\cal N}=4$ Yang-Mills theory with large $R$ charge $J$ and string theory in a pp-wave background.
\vspace{1.4cm}

\noindent
{\bf Acknowledgements}

\noindent
A.S. thanks S. Penati, A.C. Petkou and E. Sokatchev for useful discussions.\\
This work has been partially supported by INFN, MURST, and the
European Commission RTN program HPRN-CT-2000-00113 in which the authors
are associated to the University of Torino.

\end{document}